\documentclass[twocolumn]{gMOS2e}
\usepackage{subfigure}
\usepackage[usenames,dvipsnames]{xcolor}
\usepackage[pdftex,colorlinks=true,citecolor=NavyBlue,
            urlcolor=MidnightBlue,linkcolor=MidnightBlue,bookmarks=false]{hyperref}

\begin{document}
\title{Dissipative particle dynamics: Dissipative forces from atomistic simulation}
\author{
  \name{Vlad P. Sokhan and Ilian T. Todorov}
  \affil{STFC Daresbury Laboratory, Sci-Tech Daresbury, Keckwick Lane, Daresbury,
	       Cheshire WA4 4AD}}
\maketitle

\begin{abstract}
We present a novel approach of mapping dissipative particle dynamics (DPD) into classical molecular dynamics. By introducing the invariant volume element representing the swarm of atoms we show that the interactions between the emerging Brownian quasiparticles arise naturally from its geometric definition and include both conservative repulsion and dissipative drag forces. The quasiparticles, which are composed of atomistic host solvent rather than being simply immersed in it, provide a link between the atomistic and DPD levels and a practical route to extract the DPD parameters as direct statistical averages over the atomistic host system. The method thus provides the molecular foundations for the mesoscopic DPD. It is illustrated on the example of simple monatomic supercritical fluid demonstrating good agreement in thermodynamic and transport properties calculated for the atomistic system and DPD using the obtained parameters.
\end{abstract}


\section{Introduction}
As is well known, the principal workhorse of computer modelling of condensed 
matter -- molecular dynamics -- works best when all degrees of freedom are 
of equal denomination. When the problem includes both fast (local) and slow 
(collective) degrees of freedom and they are coupled, as is often the case 
in condensed matter problems \cite{Bern76}, the efficiency of the statistical 
sampling dramatically reduces. Finding efficient \emph{coarse-grained} (CG) 
solutions, i.e., based on reduced and more homogeneous sets of variables, 
is a challenging problem that has attracted much attention in recent years 
\cite{Kame11}. In general, CG leads to the generalised Brownian motion of 
the remaining variables \cite{Alle17}, which in practical terms implies a 
realisation of a stochastic process governed by a generalised Langevin 
equation (GLE) \cite{Zwan01}. Direct application of the projection operator 
approach of Zwanzig \cite{Zwan60} and Mori \cite{Mori65} to derive the 
memory kernel of the GLE is a highly nontrivial problem \cite{Hijo10} 
which can be alleviated by accepting the Markovian approximation (but 
see \cite{Cube05}). 

When the dynamics is not of primary importance and only the thermodynamic 
properties are involved, the solution is provided by a mapping of free energies 
defined in full and reduced phase spaces \cite{Voth08,Kond13}. 
Examples include polymer solutions \cite{Akke00,Akke01,Akke01a,Brie02,Alle08}, 
colloidal systems \cite{Voth08} and biological materials 
\cite{Kame11,Karp14,Levi14,Wars14}, commonly known as soft matter. 
There, a functional group of chemically connected atoms 
is typically replaced by a single bead, and several CG force fields, 
including MARTINI \cite{Marr07}, UNRES \cite{Liwo01,Liwo14} 
and SIRAH \cite{Darr15}, systematically explore this approach. 
Many methods exist for deriving CG forces from the atomistic level, including 
the force-matching method \cite{Erco94}, reverse Monte Carlo \cite{McGr88,McGr01} 
and relative entropy minimisation \cite{Shel08}.
These CG methods are able to provide accurate structure for polymers and 
biomolecular systems and have demonstrated significant speed-up due both to 
the smoothened energy landscape in the reduced configuration space 
as well as the increased timestep. With additional tuning, they are also 
capable of correctly reproducing the fluid structure. 

In contrast, deriving methods for an accurate and faithful CG dynamics appears 
to be a more complex problem for which no general solution 
exists as yet. The essence of the problem lies at the heart of CG itself. 
Simple elimination of some of the degrees of freedom (DOF) in the system takes away the associated 
channels of energy relaxation which results in accelerated transport 
properties. As a result, the CG dynamics is often an order or two faster 
than the corresponding atomistic one \cite{Lope02,Qian09,Fu13}. While 
in some cases this might be an advantage, for example by allowing fast 
determination of structure or thermodynamic properties, this precludes 
its application to fluid dynamics problems. 
In addition, as has been shown early on by D{\"u}nweg \cite{Dunw93}, 
modification of dynamics due to thermostats that violated Galilean 
invariance generally results in artificial hydrodynamic screening.

To accurately capture the hydrodynamic effects at the coarse-grained 
level, several methods have been proposed \cite{Liu16}, of which 
Dissipative Particle Dynamics (DPD) \cite{Hoog92} is particularly 
suitable for the study of dynamic and transport properties of the complex 
soft matter at the mesoscale. 
In its recent form \cite{Espa95a,Groo97} the method extends Brownian 
dynamics \cite{Alle17} to incorporate hydrodynamic effects through 
a Galilean invariant thermostat \cite{Espa95}, which consequently stipulates 
the isothermal conditions. 
The basis for this approach stems from the fact that repulsive forces 
alone can describe the structure of liquids, as has been profusely 
demonstrated in the WCA theory \cite{Week71}, and combined with a local 
short-ranged pairwise Langevin thermostat this makes DPD an attractive, 
efficient tool for multiscale interrogation of soft matter.
The advantages and limitations of the method have been discussed in a 
recent review \cite{Espa17}.

Predictive power of the method depends critically on our ability 
to relate the model parameters to a specific chemistry, to \textit{derive} 
them from first principles. Interaction between DPD particles includes 
both conservative and Langevin forces and while several methods were 
proposed for the former \cite{Erik08,Erik09,Erik09a}, the solution to 
the latter problem still eluded us \cite{Espa12}. The problem is 
particularly acute for coarse-graining the non-bonded DOF in simple 
fluids, e.g., swarms of atoms or small molecules \cite{Espa12}. 
Existing methods, depending on the thermostat used, often result in 
diffusivities that are too high (viscosities that are too low) \cite{Evan99}. 
It is also khown that the Markovian approximation employed in DPD 
equations of motion is unsound in the case of a linear force 
of the DPD ansatz \cite{Cube05}.


Here, we present a new general method of deriving the DPD model parameters, 
including dissipative forces, from atomistic simulation. In the method, we 
first define a Brownian quasiparticle (BQ) in the host atomic fluid: a CG 
object of constant volume, which both consists of fluid atoms and is immersed 
in the host fluid. The BQ is therefore an open system of fluctuating mass. 
Its motion, and more precisely its kinematics, is determined by the gross 
motion of the constituent atoms and from the analysis of its motion 
the parameters of the corresponding Brownian motion can be estimated. 
Translation of these parameters into the DPD ansatz fixes both the analytic 
shape of DPD forces and their values.

\section{Dissipative Particle Dynamics and the Brownian quasiparticles}
Dissipative particle dynamics is a mesoscopic particle-based method aimed 
primarily at hydrodynamic and rheological modelling of the fluid phase. 
In it, the atomistic nature of fluids is seen through the blurry looking 
glass of coarse-graining. As a result, despite the widespread usage of 
the method there is no precise definition of its fundamental constituent, 
the DPD particle itself. While in coarse-grained methods for bonded species 
a `group of atoms' \cite{Espa17} or molecular moiety is often used to define 
a coarse-grained bead, the situation is far less clear-cut for non-bonded 
atoms \cite{Bock07,Espa12}, where a `lump of fluid' \cite{Marsh97,Bock07} 
or a `fluid element' \cite{Fu13}, apparently 
making implicit reference to continuum hydrodynamics, is termed as a 
DPD particle.

In a sense, the DPD particle is isomorphic to the `illuminated volume' \cite{Bern76}
of light-scattering or inelastic coherent neutron scattering experiments, both 
sensitive to the local density fluctuations in the liquid. The existence of 
phonons in liquid is well established \cite{Zwan67}, and although their lifetime
is short, we want to exploit this analogy when defining the Brownian quasiparticle.
Since the DPD is not a space-filling model, the density of DPD particles is a free 
parameter and we will utilise this freedom in defining a suitable density range.

In this section our starting point is an ensemble of atoms in a certain 
thermodynamic state which we `illuminate' with a mapping ensemble of Brownian 
quasiparticles. We describe then how the interactions between Brownian 
quasiparticles and with the surrounding reservoir, mediated by exchanging 
host atoms, emerge from the definition the BQ. Molecular dynamics of the 
host subsystem drives the motion at the coarse-grained level and from the 
analysis of this motion we construct the conservative and dissipative drag 
forces which are consistent with the target DPD model.

\subsection{Mapping into atomistic system}
Despite the fact that the Brownian quasiparticle shares much in common with 
the usual Brownian particle, there is one important difference -- whereas 
a Brownian particle is just immersed in fluid, BQ is both immersed and 
\emph{consists} of fluid. As a consequence of this, the interactions 
between the BQ beads bear the birthmark of the primordial molecular level. 
Implicitly, they are distilled from a more fundamental atomistic 
representation through a coarse-graining transformation, and in order to 
distinguish them we shall call from now on fundamental level objects 
\emph{atoms} (irrespective of whether they are atoms or molecules), reserving 
the term \emph{beads} (or DPD particles) to the coarse-grained level. 
For simplicity, we shall also assume that particles are spheres, although 
they could take any convex shape as long as their volume remains invariant.

In the original papers \cite{Hoog92,Koel93} very little details were given about 
the relation between the DPD particles and atoms and they were assumed to 
represent some mesoscopic degrees of freedom, or `lumps' of fluid \cite{Marsh97}.
In order to map DPD particles into the atomistic system we propose the following 
\emph{definition}: a Brownian quasiparticle collectively represents all atoms 
contained at a particular moment of time in an invariant volume element of the 
fluid and its kinematics is defined by the collective motion of the atoms 
contained in the volume element. This definition is based on the assumption 
that the interactions between `lumps of fluid' are short-ranged; this has been 
postulated in the DPD model and although in the BQ picture indirect interactions 
extend through the host fluid leading to a `hydrodynamic tail', direct 
intervention ceases at some separation $a$ taken as the BQ size. The definition 
also implies that the mass of the quasiparticle and its `content' fluctuates 
in time according to the grand canonical conditions (open system), and that 
its velocity at a particular instant of time is given by a velocity centroid 
of all constituent atoms. The last statement, as will be shown later, 
introduces a bias in the particle number distribution and therefore in 
the chemical potential of the Brownian quasiparticle.

Using a loose analogy with gauge bosons in particle physics, we define the 
interactions between quasiparticles in terms of the exchange of `virtual particles', 
which in this case are atoms, only `virtually' present at the coarse-grained level.
There are two types of quasiparticle interactions depending on whether the 
atom leaving the volume element goes into the `reservoir' or exchanges with 
another volume element. The former describes the act of interaction with 
the `environment', or `heat bath', and contributes to Langevin's friction 
and noise, and the latter is the classical analogue of `exchange repulsion', 
which could be understood on the example of two boats on the lake 
exchanging watermelons by throwing. We note in passing that similar ideas 
were used to illustrate the origin of the shear viscosity (e.g., the `train exchange' 
example on p.~303 of Tabor's book on gases, liquids and solids\cite{Tabo91}). 
However, since our target, the DPD ansatz, does not provision for shear forces, 
the dissipative forces of DPD are related to bulk viscosity only. Also, momentum 
conservation requires that each atom could be either a member of the heat bath or 
belong to one BQ only and cannot be in two particles simultaneously. 
This condition is sufficient to devise the interactions between BQ beads. 
When BQ spheres overlap, the atoms are reassigned according to the Voronoi 
tessellation \cite{Erik09}, i.e., each atom belongs to the nearest sphere.

The above method exposes one of the intrinsic shortcomings of the DPD model, 
in which the interacting Brownian particles suspended in a solvent are modelled 
by an ensemble of interacting Brownian particles \emph{in vacuo}. In the low-density 
limit, a DPD particle, stripped of its heat bath interactions, behaves like an 
ideal classical particle, travelling ballistically, whereas a BQ of the 
described model will retain the interactions with the environment, leading to 
diffusive Brownian motion. As a result, an ideal dissipative gas (i.e., DPD 
particles without conservative forces) presents a completely different limiting 
case from the gas of ideal Brownian quasiparticles, whose interactions are 
guided by the long-range and nonlocal hydrodynamic effects in the underlying 
atomistic system.

\begin{figure}
\begin{center}
{\resizebox*{164mm}{!}{\includegraphics{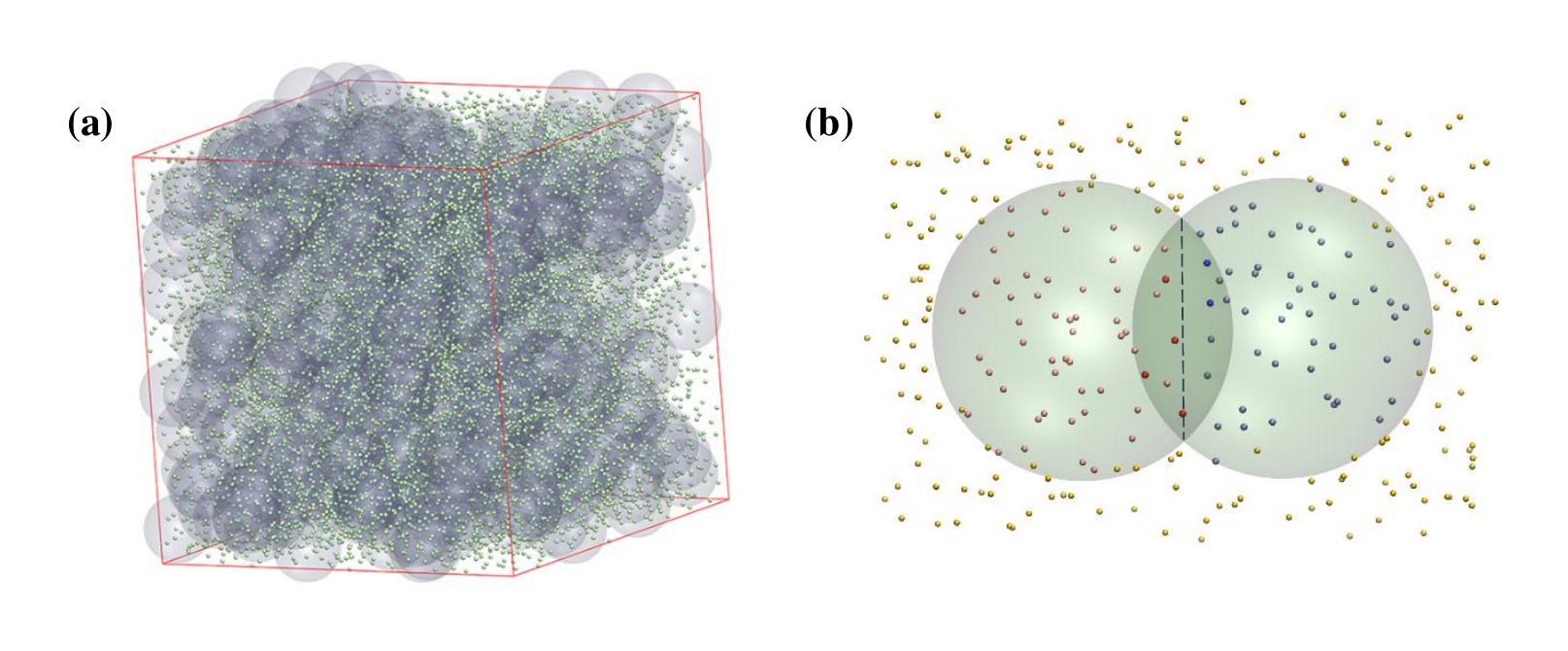}}}
\caption{(Colour online) (a) Sketch of the model. Large transparent spheres 
represent the Brownian quasiparticles, small beads -- LJ particles, and the 
system is periodically replicated in 3D. 
(b) Illustration of the two overlapping BQs. Dashed line denotes the 
position of the dividing plane separating the volumes of two quasiparticles. 
Atoms belonging to each quasiparticle are distinguished by their colour 
and in the overlap volume atoms belonging to the left BQ are red and those 
belonging to the right BQ, blue.}\label{fig_system}
\end{center}
\end{figure}

\subsection{Estimating DPD forces in molecular dynamics}
Consider an ensemble of $N$ atoms in volume $V$ prepared in initial 
state by coupling to a thermostat at temperature $T$. After initial 
equibration the system is isolated and evolved further according to 
constant energy molecular dynamics. In the production section, 
we disperse across the system 
$M$ BQ `watchers' (we are interested 
in $M<<N$) of mass $m_i,$ $(i\in[1,M])$, and size $a$, which are 
large in comparison with the corresponding atomic quantities as 
sketched in Figure~\ref{fig_system}(a). Motion of atoms is unaffected 
by the presence of coarse-grained watchers while the motion of 
the latter obeys the Langevin equations of motion \cite{Zwan01},
\begin{subequations}
\begin{equation}
\dot{\bf r}_i(t) = {\bf p}_i(t)/m_i,
\end{equation}
\begin{equation}
\dot{\bf p}_i(t) = {\bf F}_i({\bf r}_1,\ldots,{\bf r}_M,\dot{\bf r}_1,\ldots,\dot{\bf r}_M),
\end{equation}
\end{subequations}
where a dot denotes the time derivative and the force on the $i-$th particle, ${\bf F}_i$, depends on positions and velocities of all other particles in the system. 
In the pairwise approximation of DPD it is
\begin{equation}
  {\bf F}_i({\bf r}_1,\ldots,{\bf r}_M,\dot{\bf r}_1,\ldots,\dot{\bf r}_M)
	= \sum_{j\neq i} {\bf f}_{ij}(r_{ij}),
\end{equation}
where the pair force between the particles $i$ and $j$, ${\bf f}_{ij}$, acting 
along the interparticle vector ${\bf r}_{ij}\equiv{\bf r}_j - {\bf r}_i$,
consists of three collinear components: conservative, ${\bf f}_{ij}^{({\rm c})}$, 
dissipative, ${\bf f}_{ij}^{({\rm d})}$, and stochastic, ${\bf f}_{ij}^{({\rm s})}$,
\begin{equation}
 {\bf f}_{ij}(r_{ij}) = {\bf f}_{ij}^{({\rm c})}(r_{ij})
    + {\bf f}_{ij}^{({\rm d})}(r_{ij})
		+ {\bf f}_{ij}^{({\rm s})}(r_{ij}),\label{force_tot}
\end{equation}
where the last two terms, collectively defining the Langevin forces in the DPD 
ansatz, form a DPD thermostat \cite{Espa95}.
Individual force terms are written as 
\begin{subequations}
\begin{equation}
  {\bf f}_{ij}^{({\rm c})}(r_{ij}) = \alpha(r_{ij}){\bf e}_{ij},
	\label{force_c}
\end{equation}
\begin{equation}
  {\bf f}_{ij}^{({\rm d})}(r_{ij}) = -\gamma(r_{ij})u_{ij}{\bf e}_{ij},
	\label{force_d}
\end{equation}
\begin{equation}
  {\bf f}_{ij}^{({\rm s})}(r_{ij}) = \sqrt{2k_{\rm B}T\gamma(r_{ij})/\Delta t}\,\theta_{ij}{\bf e}_{ij}.
  \label{force_s}
\end{equation}
\end{subequations}
Here, we slightly modified the standard notation \cite{Espa95a,Groo97}
to accommodate more general functional forms, ${\bf e}_{ij}$ is a unit vector in the direction of the interparticle vector ${\bf r}_{ij}$, i.e., 
${\bf e}_{ij}={\bf r}_{ij}/r_{ij}$, and $r_{ij}\equiv \vert{\bf r}_{ij}\vert$ is the 
inteparticle separation.
Note that the dissipative force (\ref{force_d}) is linear in the relative 
velocity of 
two particles along the line conneting them, $u_{ij}\equiv{\bf u}_{ij}{\bf e}_{ij}$.
In the stochastic term (\ref{force_s}) $\theta_{ij}$ is a random variable, independent 
for each pair of particles, with zero mean and unit variance, i.e., 
$\langle\theta_{ij}(t)\theta_{kl}(t')\rangle = (\delta_{ij}\delta_{ij} + 
\delta_{ij}\delta_{ij}\delta(t'-t))$, where $\delta(t)$ is a Dirac delta function, and a tensor $\delta_{nm}$ is the Kronecker delta.

The model thus includes two functions, $\alpha(r_{ij})$ and $\gamma(r_{ij})$, 
whose functional form, together with the required parameters, we want to 
derive from the underlying atomistic simulation rather than postulate it 
by accepting, e.g., the standard DPD ansatz. They define conservative and 
dissipative forces, respectively, of the DPD model.

Since in both atomic and coarse-grained descriptions we are concerned with
pairwise forces only, conservative forces could be found in a simplified way
assuming the conjecture that manybody effects arise in the same way at 
both levels and we need to fix the force in the low-density limit only. 
In that case the conservative part of the interaction potential $U(r)$ 
between the Brownian quasiparticles is related to their radial distribution 
function $g(r)$ via Boltzmann inversion,
\begin{equation}
 U(r) = -k_\text{B}T\ln g(r),\label{bolt}
\end{equation}
where $k_\text{B}$ is Boltzmann's constant and $T$ is the system 
temperature. This is known to define uniquely the corresponding 
forces \cite{Hend74}. In practical terms, we first fit the calculated 
$g(r)$ to an analytic expression and then use its derivative to estimate 
the force. This route requires accurate determination of the structure 
since the accuracy of forces could be severily compromised due to 
the second step.

For the \emph{dissipative} forces a suitable procedure can be established 
by noticing that the dissipative term is the only odd term with respect
to time reversal. Therefore, by multiplying eq.~(\ref{force_tot}) 
through by ${\bf u}_{ij}$ and taking the canonical average, one obtains
\begin{equation}
 \langle{\bf f}_{ij}(r_{ij}){\bf u}_{ij}\rangle 
   = \langle{\bf f}_{ij}^{({\rm d})}(r_{ij}){\bf u}_{ij}\rangle
   = -\gamma(r_{ij})\langle{\bf u}^2_{ij}\rangle,
\end{equation}
where we took into account that both $\langle{\bf u}_{ij}\rangle=0$ and
$\langle\theta_{ij}{\bf u}_{ij}\rangle=0$.
Thus, the dissipative force can be estimated in atomistically driven 
coarse-grained simulation by taking the direct ensemble average,
\begin{equation}
  \gamma(r_{ij}) = -\frac{\langle{\bf f}_{ij}(r_{ij}){\bf u}_{ij}\rangle}
	   {\langle{\bf u}^2_{ij}\rangle}.\label{drag}
\end{equation}

The total force between two beads, ${\bf f}_{ij}$, can be directly 
estimated from its definition,
\begin{equation}
  {\bf f}_{ij}(r_{ij}) = \sum_{\substack{n\in \Omega_i,\\m\in \Omega_j}}{{\bf f}_{nm}(r_{nm})},
\end{equation}
i.e., as a sum over atomic pairs belonging to BQ beads $i$ and $j$, 
respectively. Similarly, velocity of bead $i$ is calculated as a centroid 
velocity,
\begin{equation}
  {\bf u}_{i}(t) = \sum_{n\in \Omega_i}{{\bf v}_{n}(t)}.
\end{equation}

\begin{table}
\tbl{Simulation parameters and properties of Brownian quasiparticles.}
{\begin{tabular}[l]{@{}rcccccc}\toprule
$a/\sigma$ & $M$ & $N^\text{id}_m$ & $N_m(1)$ & $N_m(0)$ & $T^*_\text{B}$ & $D^*_\text{B}$  \\
\colrule
2 & 3375 & 2.0944 & 2.098  & 2.5482 & $1.173$ & 0.075 \\
3 &  999 & 7.0686 & 7.075  & 7.4692 & $1.063$ & 0.043 \\
4 &  422 & 16.755 & 16.77  & 17.139 & $1.021$ & 0.020 \\
6 &  125 & 56.549 & 46.46  & 56.827 & $0.999$ & 0.089 \\
8 &   54 & 134.04 & 110.04 & 134.41	& $0.992$ & 0.005 \\
\botrule
\end{tabular}}
\label{params}

\end{table}

\section{Simulation results and discussion}
In order to verify and validate the method, as a test case we used, 
as a host, a simple supercritical Lennard-Jones (LJ) fluid, 
for which both thermodynamic and transport properties are well known. 
In standard notation, $\sigma$ and $\epsilon$ define the lendth and energy 
scales at this level. Single host dynamics drives the coarse-grained 
motion at various levels defined by the coarse-graining factor $N_m$, i.e.,
an average number of atoms in a BQ bead. This provides a direct way to 
calculate the distribution functions and from them the potential parameters 
for the mapping beads.

\subsection{Atomistic system}
We considered a single supercritical state of the LJ fluid at 
reduced temperature $T^*=1.8$ and reduced density $\rho*=0.5$. 
The system of 13500 particles was placed in a $30\times30\times30$ cube 
(in LJ units, see Appendix \ref{appunits}), periodically replicated in 
3D, and a set of microcanonical MD calculations was performed using the 
velocity Verlet \cite{Alle17} scheme for reduced times $t^*= 4.5\cdot 10^6$ 
using a time step $\Delta t^*=3.73\cdot10^{-3}$. 
The interactions were truncated at $2.5\sigma$, which, together with the 
small time step used, provided good energy conservation with standard relative 
deviation in energy per particle of $10^{-6}$. The system was initially 
equilibrated for $t^*=10^4$ using the Nos\'{e}--Hoover thermostat \cite{Alle17} 
with relaxation constant of $\tau^*_\text{NH}=2.3$. In subsequent production 
runs we monitored the temperature but despite the long MD integration times 
we did not notice any observable temperature drift.

All calculations were performed using a modified version of DL\_POLY \cite{Todo06}.
System pressure, calculated using the virial route and including standard 
long-range corrections for homogeneous systems \cite{Alle17}, was $p^*=0.7993$.

\begin{figure}
\begin{center}
\begin{minipage}{164mm}
{\resizebox*{8cm}{!}{\includegraphics{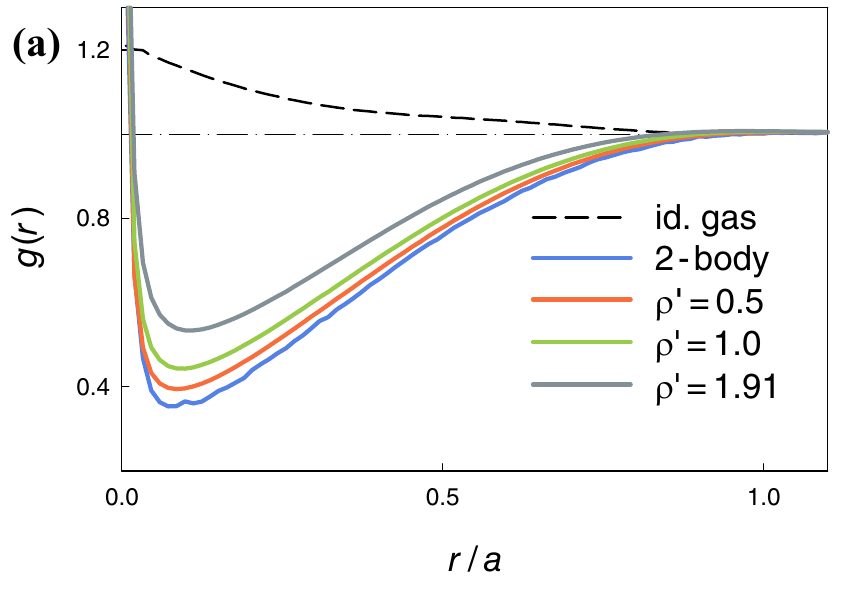}}}\hspace{6pt}
{\resizebox*{8cm}{!}{\includegraphics{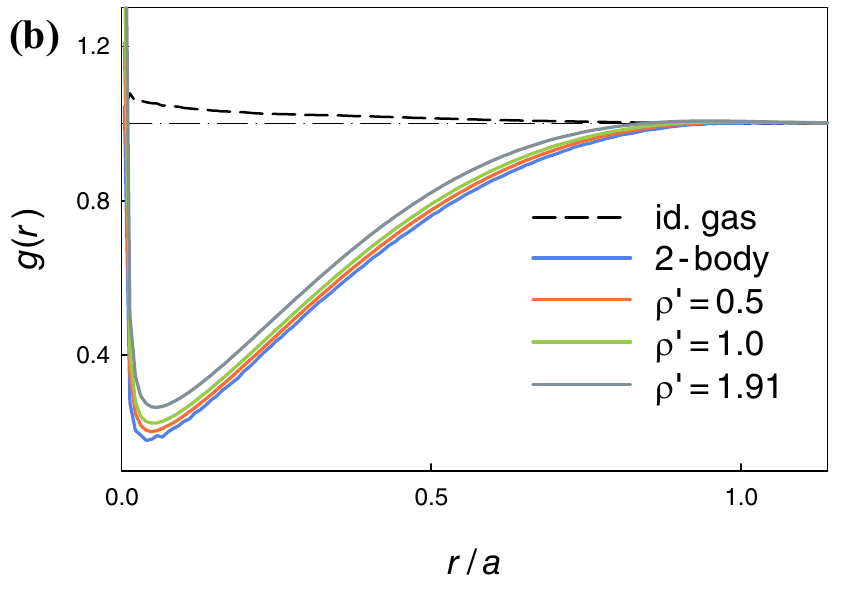}}}
\caption{(Colour online) Radial distribution functions (RDFs) for several 
values of DPD density for: (a) BQ size $a=2\sigma$, and (b) $a=3\sigma$. 
Id.~gas denotes the results for BQs that do not interact directly (see text) 
and 2-body is calculated in the low-density limit where rigorously only 
two-body forces are present.}\label{fig_rdf}
\end{minipage}
\end{center}
\end{figure}

\subsection{Brownian quasiparticles}
We calculated the structure of the BQ subsystem for a range of quasiparticle 
sizes from $a=2\sigma$ to $a=8\sigma$; they are listed in the first column of 
Table~\ref{params}.
In order to determine the appropriate density range for the matching BQ system
we recorded coarse-grained dynamics for densities (in DPD units, see 
Appendix \ref{appunits}) $\rho'=0.5$, $1.0$, and $1.908$ (the last is a 
space-filling equivalent with a volume fraction $\zeta\equiv\pi\rho'/6=1$ 
\cite{Hoog92}). The number of BQ beads used in simulation depends both on the 
coarse-graining ratio and the density of BQ. The second column of Table~\ref{params} 
gives these values for $\rho'=1.0$, and for other densities the 
numbers scale accordingly. From the definition of the interbead interactions 
it is clear that the coarse-graining ratio depends on density, and columns 
3 to 5 of the Table give values for the `ideal' coarse-graining ratio, 
$N^\text{id}_m=\rho^*a^{*3}$, a purely geometric quantity which is 
equivalent to averaging over beads randomly distributed through the system, 
as well as the values for density $\rho^*=1$, $N_m(1)$, and for an ensemble 
of diffusing quasiparticles that do not `see' each other and interact only 
with the host system, $N_m(0)$. Note that in the last case quasiparticles 
effectively interact via hydrodynamic fluctuation forces \cite{Ivlev01} 
resulting in short-range spatial correlations, as evidenced by their radial 
distribution funtions (RDF) shown by dashed lines in Figure \ref{fig_rdf}. 
As a result, the values for $N_m(0)$ are always higher than the corresponding 
ideal values although the effect decreases for larger values of $a$. 

To verify that the system-size effects do not introduce any bias in 
calculated properties at the highest density (volume fraction 1) for 
each value of $a$ we repeated the calculations using a smaller host system 
containing 4000 atoms in a cubic cell of dimensions $20\times20\times20$ 
with proportionally reduced number of quasiparticles. We found no observable 
differences in the BQ structure.

Figure~\ref{fig_rdf} presents the results obtained for the BQ radial 
distribution function calculated for several densities for two smaller 
BQ sizes: (a), $a=2\sigma$ and (b), $a=3\sigma$. RDF density dependence, 
which is a manifestation of manybody effects, is weaker for larger 
quasiparticles. When the coarse-grained subsystem consists of two 
particles only, their radial distribution function, as follows from 
the Henderson's theorem \cite{Hend74}, is uniquely defined from their 
interaction. In this case the potential (up to an irrelevant constant) 
can be directly obtained from Boltzmann inversion, 
Equation~(\ref{bolt}). We used this route to calculate the conservative 
interactions where, in order to improve sampling efficiency in this 
low-density case, we used a noninteracting `ensemble' of 1000 BQ pairs.

\begin{figure}
\begin{center}
{\resizebox*{8cm}{!}{\includegraphics{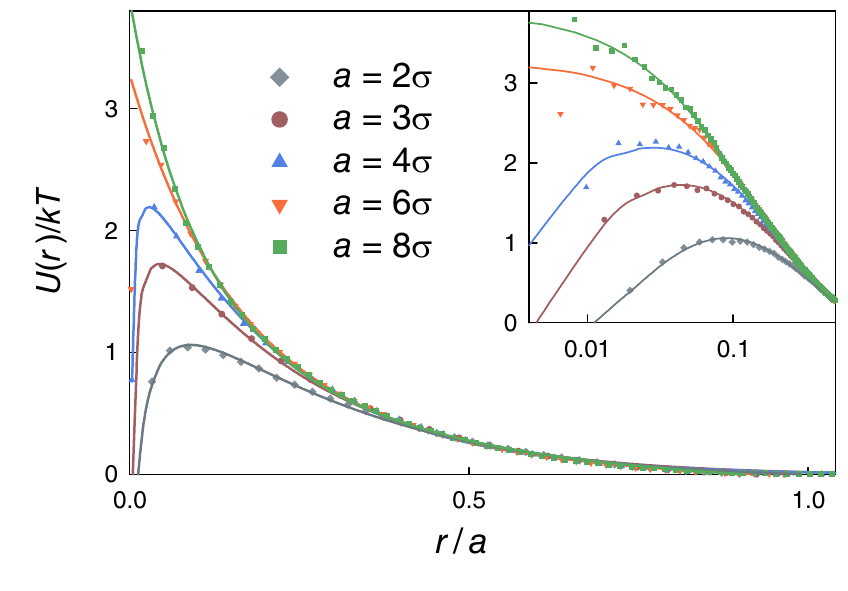}}}
\caption{(Colour online) Potential of mean force for DPD particles of several dimensions. Symbols -- simulation results, lines -- truncated exponential 
fit using Equation (\ref{pmf_fit}). The inset shows initial part of the 
distributions at low BQ separations, note the log scale for the 
separation.}\label{fig_pot}
\end{center}
\end{figure}

The conservative potential calculated for all considered BQ sizes is shown in 
Figure \ref{fig_pot}. The inset illustrates the damping of repulsion at short 
range due to correlated motion of the particles at this range leading, for BQ 
of smaller sizes, to attraction at extremely short distances. At two larger 
coarse-graining ratios corresponding to $a=6\sigma$ and $a=8\sigma$ the 
conservative potential decreases monotonously.
Solid lines are the results of the fit using Equation (\ref{pmf_fit}).
The obtained parameters of conservative potential and force for three smaller
quasiparticles are given in Table~\ref{tab_par}.

\begin{figure}
\begin{center}
\begin{minipage}{164mm}
{\resizebox*{8cm}{!}{\includegraphics{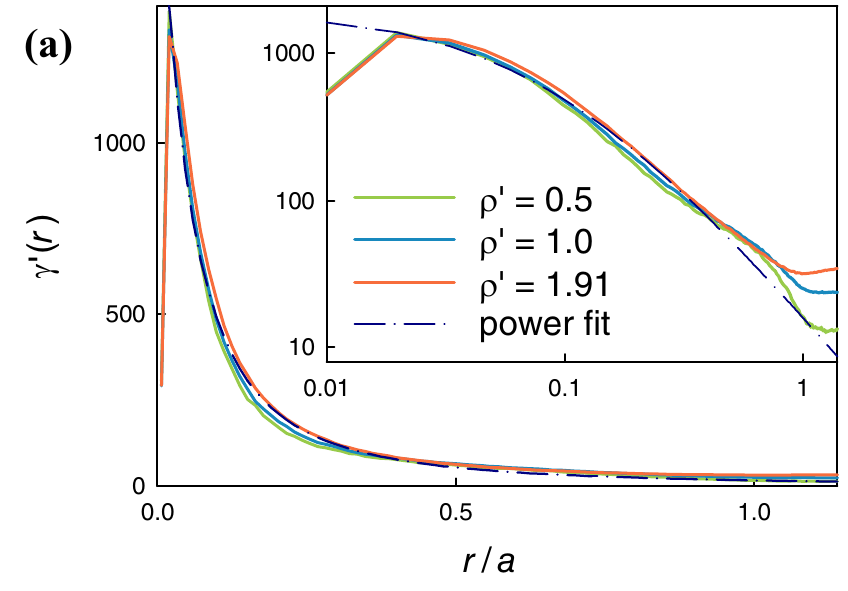}}}\hspace{6pt}
{\resizebox*{8cm}{!}{\includegraphics{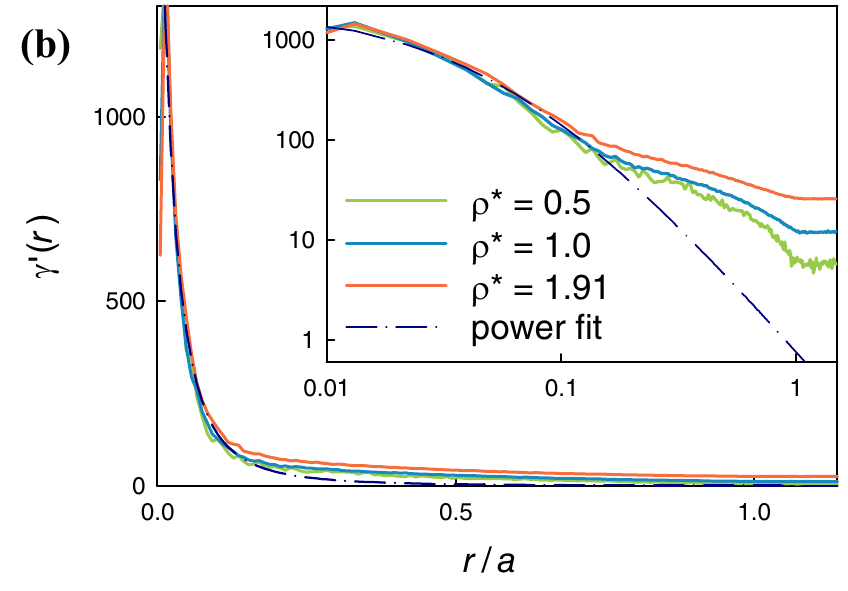}}}
\caption{(Colour online) Viscous drag foefficient for several values of BQ 
         density for $a=2\sigma$ (a) and $a=3\sigma$ (b). Full lines, 
				 simulation results; dashed lines, power fit using Equation \ref{fit_dis}.
				 Insets show the corresponding log-log plots.}\label{fig_drag}
\end{minipage}
\end{center}
\end{figure}

In order to obtain the parameters of the dissipative forces, we calculated 
the distance dependent viscous drag coefficient $\gamma'(r)$ using 
Equation (\ref{drag}). The results, present in Figure \ref{fig_drag} 
for several studied densities for two smaller BQ sizes, $a=2\sigma$ (a) and 
$a=3\sigma$ (b), almost collapsed to a single line in the former case, 
where the density dependence occurs only at separations greater than the BQ 
size. For larger coarse-graining ratio the density-dependent hydrodynamic 
effects are however visible in the BQ overlap region, $a'<1$.
The parameters of the fits for three smaller BQ cases are given in 
Table \ref{tab_par}.

Finally, the velocity autocorrelation functions (VACF) of BQ particles, 
$C(t)$, provide a link to transport properties of the coarse-grained subsystem.
From the energy equipartition it follows that 
\begin{equation}
  C(0)=k_\text{B}T^*_\text{B}/\bar{M^*_a}, \label{equip}
\end{equation}
where $T^*_\text{B}$ is the temperature of BQ subsystem (which can deviate
from the temperature $T^*$ ($T'=1$ in DPD units) of the atomic host) and 
$\bar{M^*_a}$ is the mean mass of a BQ particle. It is related to the 
coarse-grain ratio, i.e., the mean number of atoms in a BQ, via 
$\bar{M^*_a}=N_m(\rho')$ (recall that $m^*=1$). From (\ref{equip}) we 
estimate the temperature of the BQ subsystem $T^*_\text{B}$ and the 
values of $T'_\text{B}\equiv T^*_\text{B}/T^*$ for $\rho'=1$ are given in 
Table \ref{params}. It occurs that the deviation from $1$ at lower 
values of $a$ is due to the presence of BQs with zero atoms.
When for $a'=2$ these cases were excluded from the average, temperature dropped to $1.01$.

\begin{figure}
\begin{center}
\begin{minipage}{164mm}
{\resizebox*{8cm}{!}{\includegraphics{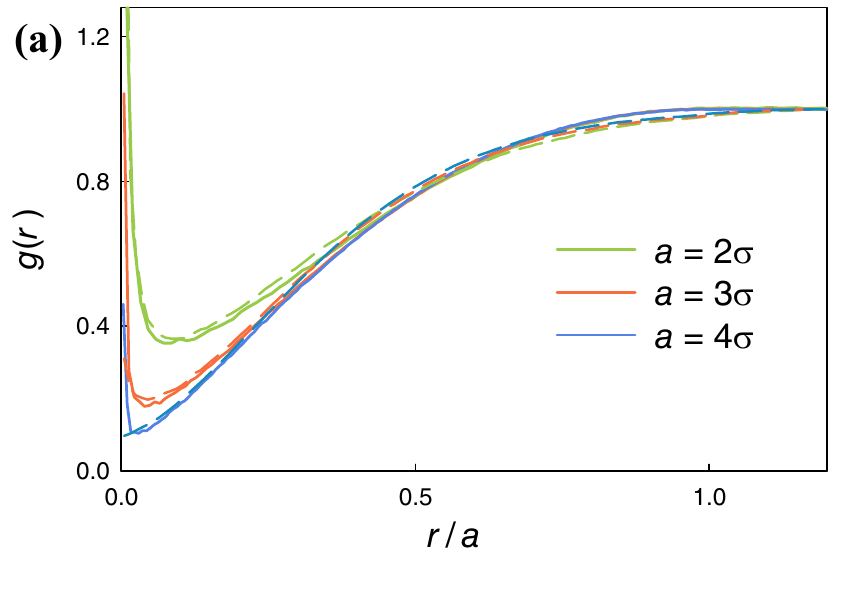}}}\hspace{6pt}
{\resizebox*{8cm}{!}{\includegraphics{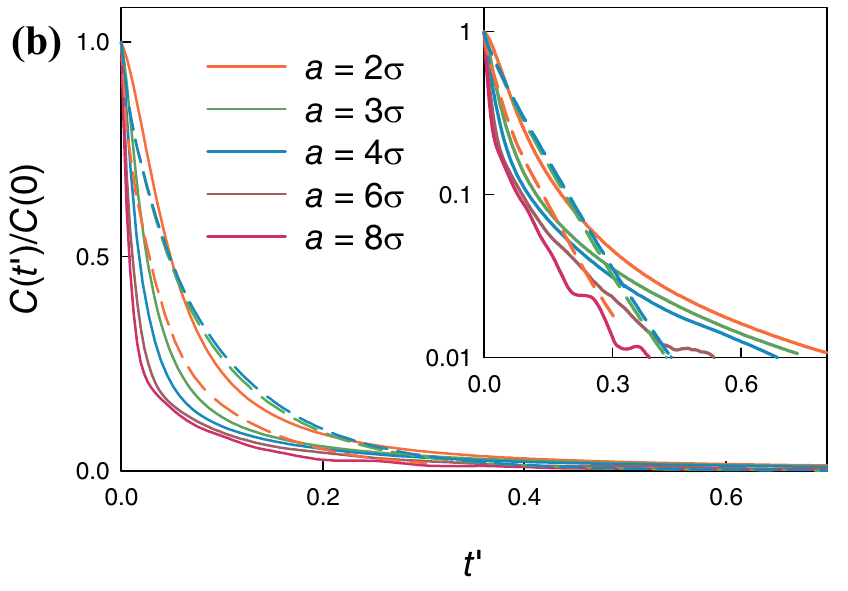}}}
\caption{(Colour online) (a) RDFs calculated for the matching BQ (full lines) 
and DPD (dashed lines) systems for three coarde-graining ratios defined by $a$.
(b) Velocity autocorrelation function of DPD particles, averaged over $C_{xx}$, 
$C_{yy}$, and $C_{zz}$, for $N = 36$ and $a = 6\sigma$. The inset is the log--log 
plot of the same function to illustrate the long-range behaviour. The grey dashed 
line is the best linear fit for these coordinates at large times.}\label{fig_vacf}
\end{minipage}
\end{center}
\end{figure}

\subsection{Coarse-grained level}
In order to test and validate the method we performed a series of DPD 
simulations using the software package DL\_MESO \cite{Seat13}, which 
was modified to include the new potential forms for conservative and 
dissipative forces and to calculate on-the-fly radial distribution and 
velocity autocorrelation functions.
In principle, any molecular dynamics integration scheme can be adapted to 
integrate the DPD SDE \cite{Goga12}. In practice, however, the velocity 
Verlet (VV) scheme \cite{Groo97} is often the standard choice. The 
presence of velocity-dependent Langevin forces require modification of 
the VV scheme \cite{Vatt02}, and in our calculations we used DPD-VV 
scheme as implemented in the DL\_MESO package.

Using analytic expressions for the conservative and dissipative forces 
defined in the appendix by (\ref{fit_con}) and (\ref{fit_dis}), correspondingly, 
and parameters given in Table \ref{fits}, we performed a series of validating 
DPD calculations. We used a system containing 1728 DPD beads in a cubic box with 
the side length of $L=12a$, which corresponds to a reduced density $\rho'=1.0$.
To check the accuracy of the conservative forces, we calculated the radial 
distribution functions of DPD beads and compared them with the reference data 
obtained for the Brownian quasiparticles (Figure~\ref{fig_vacf}a). The differences 
between the DPD and reference $g(r)$ are comparable with the inaccuracies of the 
fit, and are smaller than the variation in RDFs due to density changes. This 
confirms our assumption about the transferability of manybody effects.

Comparison of the DPD velocity autocorrelation functions with BQ for all coarse-graining
values shown in Figure~\ref{fig_vacf}b illustrate the differences in these 
approaches and highlight the problems of the DPD model. All DPD results are 
missing the characteristic features of atomistic VACF, which are present in BQ 
VACF: zero slope at the origin and the hydrodynamic long-range tail. 
Fitting BQ VACF to an inverse power gave the exponent between 2 and 3 (the last 
column in Table \ref{tab_par}) whereas for DPD the VACF decays exponentially.
However, the values of VACF for BQ and DPD are of the same order of magnitude and 
with a better fit the agreement could be improved.

\section{Conclusions}
We have presented a new approach towards fundamental derivation of model parameters 
for coarse-graining the nonbonded degrees of freedom within the dissipative particle 
dynamics framework. Translation of atomistic dynamics onto coarse-grained 
dissipative dynamics is a complex process and in our approach we used two stages 
which reflect its different strands and which require separate assumptions.
In the first stage, we introduced a Brownian quasiparticle as an open Lagrangian 
system that could be of any shape. For convenience we considered spherical 
quasiparticles only but the aspherical case is required to describe the rotational 
diffusion. The atomic flux in and out of the quasiparticle defines the 
dissipative and stochastic forces acting on it and the condition that any atom 
at any time could be a part of at maximum one quasiparticle defines the 
conservative forces. 
The atomistic host system drives the dynamics of quasiparticles through the 
interaction of atoms constituting quasiparticles with the `environment', 
i.e., the rest of the atomistic system.
At this stage the mapping is straightforward and the motion of quasiparticles 
gives a faithful description of the fluid motion at the coarse-grained 
level defined by the dimensions of quasiparticles.

The second stage involves recasting the Brownian quasiparticle dynamics in 
the form of DPD. This includes several approximations required by the 
DPD ansatz, including pairwise approximation for all forces, Markovian 
approximation for Langevin forces \cite{Zwan01} and truncation of viscous 
drag forces at the DPD particle size.

As shown in this work, the proposed method provides a working route to 
derive the DPD parameters from the underlying atomistic system giving a 
quantitative agreement of the structure and qualitative description of the
dynamics. The proposed concept of Brownian quasiparticle is found to be 
constructive in revealing the links between the host and coarse-grained 
description. Focusing on describing the method rather than exploring all 
its capabilities, we omitted several important quations that need to be 
investigated further. Thus, an interesting question is whether, in the 
dependence of coarse-grained dynamics on the bead size, there are any 
magic numbers. This would be the case if the BQ describe the collective 
variables in fluid dynamics.

\section*{Acknowledgements}
All calculations were performed on Hartree Centre's computers.
VPS acknowledges useful discussions with Patrick Warren and Dominic Tildesley.

\section*{Funding}
This work was supported in part by funding from the CCP5 Programme.

\bibliographystyle{gMOS}
\bibliography{DPD_refs}
\appendices
\section{A note about the systems of units in DPD modelling}\label{appunits}
We used two systems of units defined by the relevant subset of base units 
that includes length (dimension symbol L), mass (dimension symbol M), and 
time (dimension symbol T). On the atomistic scale it is the Lennard-Jones (LJ)
system of units, denoted by an asterisk. They are sometimes called dimensionless 
or reduced units. In terms of LJ paricle size, $\sigma$, potential well depth, 
$\epsilon$, and unified atomic mass unit, $u$, relevant quantities are
length $L^*\to L/\sigma$, energy $E^*\to E/\epsilon$, force $F^*\to F\sigma/\epsilon$,
viscous drag $\gamma^*\to\gamma\sigma/\sqrt{\epsilon m}$, and diffusivity 
$D^*\to D\sqrt{m/\epsilon}/\sigma$. 
On the coarse-grained level, we used a DPD system of units, where a DPD particle
size, $s$, defines the length scale and kinetic energy, $kT$, specifies the energy scale.

\section{Details of the fit}\label{fits}
\begin{table}
\tbl{Parameters of the fit functions for conservative and dissipative forces.}
{\begin{tabular}[l]{@{}lccccccc}\toprule
$a$ & $A$ & $r_0$ & $c$ & $d_0$ & $\gamma(0)$ & $r_\text{d}$ & $\beta$ \\ \colrule
2$\sigma$ & 2.354 & 0.286 & 0.0730 & $0.0210$ &  1940 & 0.0932 & 2.00 \\
3$\sigma$ & 2.870 & 0.247 & 0.0601 & $0.0407$ &  5469 & 0.0100 & 1.50 \\
4$\sigma$ & 2.985 & 0.225 & 0.0361 & $0.0625$ & 14773 & 0.0326 & 2.35 \\
\botrule
\end{tabular}}
\label{tab_par}
\end{table}

For the Brownin quasiparticles of diameter $s$ we fit the conservative potential to a truncated exponential
\begin{equation}
 U(r) = A\exp(-r/r_0) - c/(r+d_0),\quad(r\leq s)\label{pmf_fit}
\end{equation}
with the parameters of the fit given in Table~\ref{tab_par}. This gives rise to 
the conservative force (\ref{force_c}) with
\begin{equation}
 \alpha(r) = A\exp(-r/r_0)/r_0 - c/(r+d_0)^2.\label{fit_con}
\end{equation}
For the dissipative force (\ref{force_d}), we fit the viscous drag to an 
inverse power
\begin{equation}
 \gamma(r) = \gamma(0)/(1+r/r_\text{d})^{\beta}.\label{fit_dis}
\end{equation}

\end{document}